# Numerical Reconstruction of 2D Magnetic Focusing Experiments

Dongsung T. Park [1*], Seokyeong Lee[1*], Uhjin Kim [2], Yunchul Chung [3], Hyoungsoon Choi [1†], Hyung Kook Choi [2†]

[1] *Department of Physics, KAIST, Daejeon 34141, Republic of Korea*

[2] *Department of Physics, Jeonbuk National University, Jeonju 54896 Republic of, Korea*

[3] *Department of Physics, Pusan National University, Busan 46241, Republic of Korea*

[†] h.choi@kaist.ac.kr

[†] hkchoi@jbnu.ac.kr

**Spatial aspects in quantum mechanics are often difficult to model in geometrically intricate settings that are typical of mesoscopic physics. In such cases, predicting the device behaviors is a vital but difficult challenge. Transverse magnetic focusing (TMF) is a prime example where a classically simple effect becomes difficult to approach in the quantum regime. Here, we have simulated a realistic TMF device and compared the results to those from experiments performed on GaAs/AlGaAs two-dimensional electron gas systems. Unlike previous studies, device features such as quantum point contacts and disorder were realized within the simulation. The simulated and experimental focusing spectra showed good agreement, and the analysis was extended to multichannel and energy-modulated scenarios. By revisiting the energy-modulated simulation with a quantum dot (QD) emitter, we confirmed that the unique geometry of a QD does not affect the focusing spectra, thereby validating the feasibility of such experiments in the study of monoenergetic excitations.**

Introduction

Spatial manifestations of the wave-particle duality illuminate some of the most striking features of quantum mechanics [1–7]. However, experimental investigations typically remain elusive, and theoretical studies are often hampered by the difficulty in solving bulky scattering problems. This rings particularly true in mesoscopic physics where, unlike in bulk systems, every detail matters and device geometries are becoming increasingly complex [8–10]. Consider, for example, the transverse magnetic focusing (TMF) effect [7,11–14]. The classical version is deceivingly simple; it is merely a geometric variation of the cyclotron problem that can be visualized with trivial difficulties. However, the quantum version in condensed matter is much more nuanced due to the wave nature of particles, generally affected by the awkward boundary conditions and the band structures [14–21]. It is a prime example where a classically simple phenomenon becomes highly convoluted, where even simulations remain inaccessible to those without the knowledge of scattering theory and numerical methods.

Herein, we present the simulation of a mesoscopic transverse magnetic focusing device on a typical two-dimensional electron gas system using the python package KWANT [22]. Contrary to previous numerical TMF studies [16,23], we have realized experimental features such as the quantum point contacts (QPC) used to emit and collect the electrons as part of the simulation. The simulated TMF results showed good agreement with our experimental data, and we explained the experimental discrepancies from an ideal TMF spectrum by introducing a random background potential. From the simulation, we visualized the current densities in order to verify that the wave-particles indeed perform cyclotron motion. The results were then extended to energy-modulated scenarios, which approximately corresponds to non-equilibrium ballistic transport in Fermi gas models. After analyzing TMF with varying energy levels, the QPC-emitter was replaced with a Quantum Dot (QD). The simulated results implied that the use of a different emitter led to no discernable differences within the Fermi-gas model.

**Methods**

**Experiment.** Typical mesoscopic transport experiments were performed on a device fabricated on a GaAs/AlGaAs heterostructure. A two-dimensional electron gas (2DEG) with an electron density of $n = 2.3 \times 10^{11}$ cm$^{-2}$ and mobility of $\mu = 3.8 \times 10^6$ cm$^2$/Vs resided 75 nm below the cap surface. Using an effective mass value of $m^* = 0.067\, m_e$, where $m_e$ is the bare electron mass, the electron density corresponds to a Fermi energy of $E_f = 8.2$ meV and, equivalently, a Fermi wavelength of $\lambda_f = 52$ nm. Using standard electron beam lithography, metal Schottky gates 75nm wide were deposited on the cap layer. Placing a negative voltage on the gates depletes the mobile electrons beneath the gates, allowing us to define QPCs in the 2DEG layer. The transport properties of the device were obtained using the typical lock-in technique. The conductances $G$ were analyzed by their transmission coefficient $T = G/G_0$ where $G_0 = 2e^2/h$ is the conductance quanta.

**Simulation.** A scattering center was defined using a spin-less square lattice, and the lattice properties were set by the onsite potential $U$ and hopping parameters $t$ as per the usual tight-binding formulation [24]. The Schottky gates were simulated by imposing the gates' electrostatic potentials $\phi$ onto the ungated onsite potential, i.e. $U = 4|t| + \phi$ where $|t| = \hbar/2m^*a^2$ for the reduced Planck constant $\hbar$, effective mass $m^*$, and lattice constant $a$. The gates' potential was calculated using the pinned-potential boundary condition, effectively elevated 50 nm from the lattice [25]. The magnetic field was applied through the Peierls substitution for a linear gauge symmetric in the x-direction. The particles were assigned a negative charge of $-e$, the charge of an electron. Although the position and energy units are arbitrary, here we have defined them to correspond to nm and meV as is commonly used in experiments. A 2000 nm × 2000 nm space was spanned by a square lattice with a lattice constant of $a = 5$ nm and an effective mass $m^* = 0.067 \times m_e$ with massive Fermi gas systems like our 2DEG in mind. The energy level $E_f^* = 7$ meV was used as a reference 'Fermi energy'. KWANT calculates the single-particle scattering matrix between the leads, corresponding to the conductance measurements in our experiments. The scattering matrix was used to calculate the conductance $G$ expressed by the corresponding transmission $T = G/G_0^*$ where $G_0^* = e^2/h$. Note that $G_0 = 2G_0^*$ due to the absence of spin in our simulation.

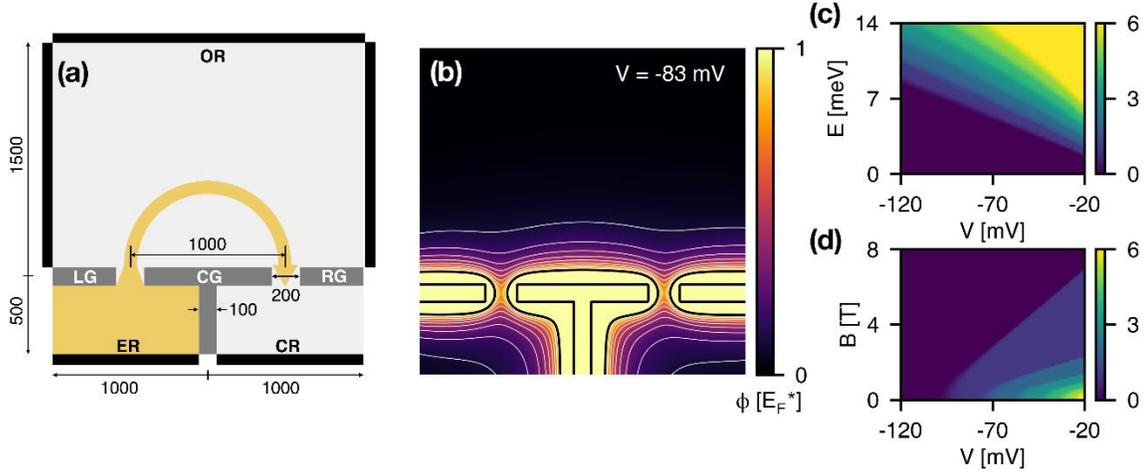

**Figure 1 Simulated Device Characteristics.** (a) Schematic cartoon of the device. The TMF device was simulated by a scattering center consisting of a 2000 × 2000 nm region with three gates (left, center and right; LG, CG, and RG) which divided the simulated area into three reservoir regions (emitter, open, and collector; ER, OR, and CR). The space between LG and CG formed the emitter QPC; CG and RG, the collector. Leads were attached to the boundary of each reservoirs (black lines); OR had three such leads on all sides to minimize the effect of currents reflected off the sides. (b) The potential map of the device when both QPCs were at the center of their first conductance plateau (all gates at a voltage $V = -83$ mV). The rectangular black line denotes the gates' outline, and the rounded curves denote where depletion would occur, i.e. $\phi = E_f^*$. Each white curve denotes a equipotential line with spacing $0.1 \times E_f^*$. The QPC was characterized by their transmissions for varying $V$ and either (c) the energy $E$ or (d) the magnetic field $B$.

## Results

**Simulated Device Characteristics.** Transverse magnetic focusing is an open system phenomenon where the Lorentz force refocuses a collimated beam of free charged particles onto the point across its cyclotron orbit. In mesoscopic physics, a beam of electrons (or holes) is typically channeled into an open reservoir using an emitter QPC [11]. A collector QPC is placed at a distance $L$ away from the emitter, perpendicular to the beam's direction, and accepts the incident charge carriers. In a 'two-point probe' scheme, the collected charges are drained at the collector reservoir while the rest (reflected) charges are drained by the open reservoir.

In TMF simulation, the QPC sources are often modeled as a point source at the boundary of the simulated space [16,23,26–29]. Here, we formed the QPC within the simulation using the dimensions from our experimental device. Figure 1(a) show the schematic presentation of the device. The device consists of three gates—left, right, and center gates (LG, RG, and CG)—which divide the 2DEG into three reservoirs—emitter, open, and collector reservoirs (ER, OR, and CR). The LG and CG form the emitter QPC; the CG and RG, the collector. As an appropriate out-of-plane magnetic field $B$ is applied, particles channeled from the ER to the OR enter their cyclotron orbits and approach the collector. The beam enters the CR through the QPC when the cyclotron radius equals half of the emitter-collector distance (Fig.1(a), yellow), i.e.

$$B_0 = \frac{p}{eL/2} \tag{1}$$

where $p$ is the beam particle's kinetic momentum. At multiples of this focusing field $B_0$, the particles perform half of their cyclotron orbits then reflect off of the CG, repeating until they eventually reach the collector. These extended trajectories are called the skipping orbits. The particles which are not collected exit the scattering center via the leads' defined on the open reservoir. Note that the open reservoir leads are defined on three edges (Fig. 1(a), red lines) in order to minimize the effect of particles reflected off the side boundaries. The simulated device parameters were chosen to resemble the dimension of the experimental device parameters (Supplementary Fig. S1).

Figure 1(b) is a topographic plot of the gate potential where all gates are imposed with the same 'voltage' $V = V_{LG}, V_{CG}, V_{RG}$. The equipotential line at which the gate potential equals $E_f^*$ (Fig. 1(b), black line) gives us the width of the QPC at its narrowest point, 76 nm, which results in the uncertainty of the focusing distance margin by 7.6 %. The QPCs were characterized with respect to two other parameters, the particle energy $E$ and the out-of-plane magnetic field $B$. Figure 1(c) plots the QPC transmission against $E$ and $V$; as expected, $V$ at which the QPC closes decreases as $E$ increases. This simply reflects the fact that particles with higher energies are blocked only by higher barrier potentials [30]. Figure 1(d) plots the QPC transmission against $E$ and $B$; $V$ at which the QPC closes increases as $B$ increases. This occurs for two reasons: a higher magnetic field has the effect of increasing the particle's effective mass in the confined, lateral direction [31]; and the magnetic flux threading the unit lattice area is much larger in the simulation than in the experiments, leading to band transformation as seen in Hofstadter's butterfly [32]. The latter effect can be minimized by decreasing the lattice constant, but such actions are not necessary for low fields of our interest $B \ll B^*$ where $B^* = 165$ T is the field at which one flux quantum $h/e$ threads a unit cell area $a^2$. The familiar characteristics seen in Figs. 1(c, d) illustrate that the QPCs defined in the simulation are faithful representations of their experimental counterparts.

**TMF & Current Density.** The focusing spectrum corresponds to the transmission or current from the emitter to the collector QPC as the magnetic field is varied. Figure 2(a) is the simulated focusing spectrum. Inspecting the system's current density illustrates the main features of the spectra. Initially, the emitted current is collimated to follow a straight line (Fig. 2(b)). As the magnetic field is introduced, the Lorentz force deflects the current (Fig. 2(c)) until the particle impinges upon the collector (Fig. 2(d)), leading to a transmission peak in the focusing spectra. In the simulation, the focusing length was set to $L^* = 1 \: \mu m$ which corresponds to a focusing peak at $B_0^* = 146$ mT. The peak was actually observed at 158 mT giving a deviation from the prediction by 8 %, which is acceptable considering the QPC width affecting the focusing distance. At nearly double the observed focusing field, 333 mT, the first skipping orbit peak is observed and the corresponding trajectory can be seen from the current density as well (Fig. 2(e)). At 998 mT (Fig. 2(a), dashed line), the cyclotron orbit is smaller than the QPC width (76 nm) and no current should be reflected off the collector. This effect is seen as a plateaued, unitary transmission, $G = e^2/h$, from the emitter to the collector in the focusing spectra (Fig. 2(a), purple circle). The corresponding current density resembles quantum Hall edge channels as shown in Fig. 2(f) [23]. At yet higher fields, the QPC closes (Fig. 2(g)).

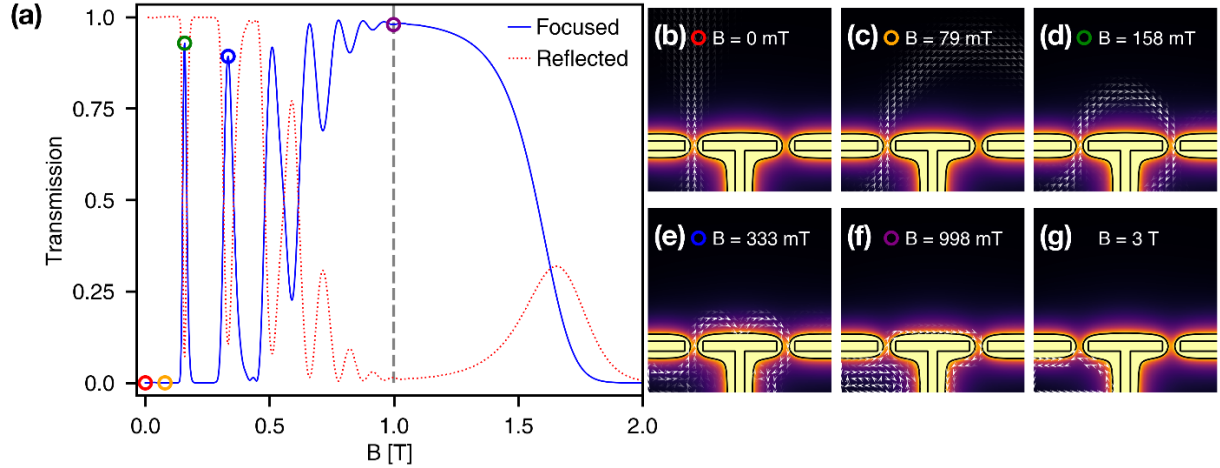

**Figure 2 TMF & Current Density.** (a) The transmission of currents from the emitter focused to the collector (blue) or reflected off to the open reservoir (red). (b) Without a magnetic field, the current headed straight after leaving the emitter. (c) The presence of a magnetic field incurs a Lorentz force which bended the current until (d) the current was focused onto the collector. (e) At certain higher fields, the current were focused after skipping off a central barrier. (f) The trend continued until the cyclotron diameter was smaller than the QPC width, i.e. $B \geq 998$ mT ((a), dashed line), at which point the classical trajectories dictates that all current be totally transmitted into the collector. (g) The quantum Hall edge-like current was blocked off when the magnetic field is strong enough raise the lowest QPC subband above $E_f^*$.

**Experiment & Disorder.** The focusing spectrum obtained from our experiment closely resembled the simulation results with a few caveats. The experiment was done using a focusing length of $L = 1.5$ μm, corresponding to a focusing field of $B_0 = 100$ mT (Fig. 3(a)). While the focusing peaks were observed, neither were the peaks unitary nor did the collected current form a clean plateau at high magnetic fields. We were able to reproduce such behaviors in our simulation by adding a disorder potential. The disorder was modeled using a random background potential $\Phi(x)$, i.e. $\phi[\Phi] = \phi + \Phi$. In experiments, such potentials are unavoidable due to the ionized dopant layer and lattice imperfections [33–36]. In simulation, the random potential was created by gaussian smoothing a normally distributed field (Supplementary Fig. S2, S3). The smoothing was parametrized by the kernel's width $\sigma_l$ and the potential's standard deviation $\sigma_\Phi$, i.e. $\overline{\Phi(x)\Phi(0)} = \sigma_\Phi^2 \exp(-x^2/2\sigma_l^2)$.

The disorder potential was analyzed for 9 combinations of $\sigma_\Phi = 0.1$, 0.3, and 1 meV at $\sigma_l = 10$, 30, and 100 nm. For small disorders (Figs. 3(b-d), $\sigma_\Phi = 0.1$ meV), the focusing spectrum showed little change at low magnetic fields. However, the unitary transmission plateau after $B > 1$T from Fig. 2(a) started to exhibit unpredictable oscillations, likely due to the local scattering events at the rough gated boundaries leading to quantum interferences. At $\sigma_\Phi = 0.3$ meV (Figs. 3(e-g)), the disorder's effect became evident for $\sigma_l = 10$ and 30 nm as the focusing peaks were suppressed and the unpredictable oscillations at higher fields grew to a sizeable fraction of the total transmission. For $\sigma_l = 100$ nm, the disorder did not affect the spectrum as much. In particular, the oscillations seen for $\sigma_l = 30$ nm (fig. 3(f)) were comparable to those seen in the experiment and the gated boundary becomes visibly irregular (fig. 3(k)). The rough boundary reflects the particles in an unpredictable manner, and the trajectories become increasingly complex with each skipping orbit. The erratic transmission oscillations at higher magnetic fields can be attributed to the interference between such irregular paths. When $\sigma_\Phi = 1$ meV (figs. 3(h-j)), the current is almost entirely lost due to the strong disorder, likely due to localization effects [37]. From this analysis, we conclude that aforementioned non-ideal properties in our experiments can be accounted for by introducing a random potential on the order of $\sigma_\Phi = 0.3$ meV and $\sigma_l = 30$ nm.

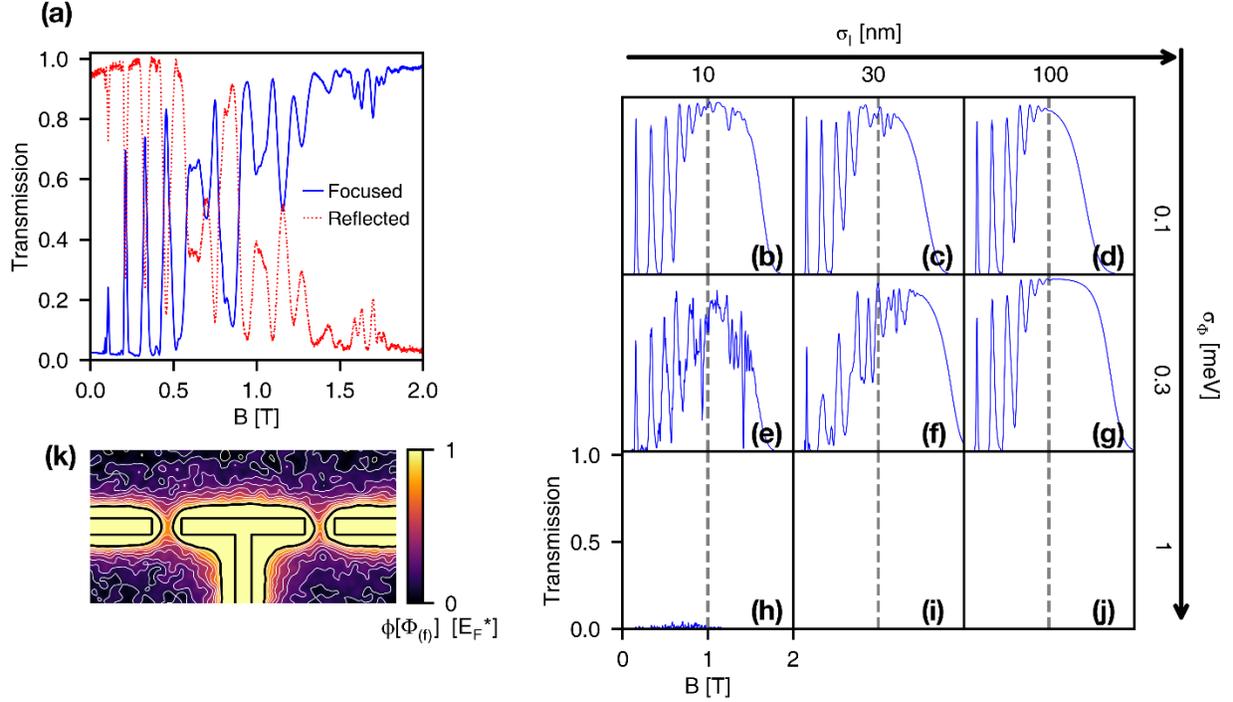

**Figure 3 Experiment & Disorder.** (a) The TMF experiment presented a focusing spectrum similar to the simulated results. However, the peaks were much smaller at times, and the transmission to the collector exhibited irregular oscillations before it becomes unitary at high $B$ fields. (b-j) The phenomenon was modeled using a disorder potential in the background for various values of disorder height $\sigma_\Phi$ and correlation length $\sigma_l$. For weak disorders $\sigma_\Phi = 0.1$ meV, the unpredicted oscillations occurred at an amplitude smaller than those seen in the experiments but the peak heights were unaffected. For moderate disorders $\sigma_\Phi = 0.3$ meV, oscillations had a typical amplitude similar to those seen in experiments and the peak heights were suppressed as well. For strong disorders $\sigma_\Phi = 1$ meV, no current was transmitted. At small correlation lengths $\sigma_l = 10$ nm, the oscillations occurred much more sharply than those seen in experiments. At moderate correlation lengths $\sigma_l = 30$ nm, the oscillations had a width comparable to those seen in experiments. At large correlation lengths $\sigma_l = 100$ nm, such oscillations disappeared. The disorder potential at $\sigma_l = 30$ nm and $\sigma_\Phi = 0.3$ exhibited properties most similar to the experiment; the potential landscape is plotted in (k)—each white curve denotes an equipotential line with spacing $0.1 \times E_f^*$.

**Multichannel TMF.** The simulation can be easily applied to scenarios where the QPC channels are greater than one. Figure 4(a) is the TMF spectra for obtained by setting the collector QPC to two channels and varying the injector QPC width via $V_{LG}$. At first glance, the result resembles the direct product of a focusing spectrum and the QPC magnetoconductance. The focused transmission oscillates with the magnetic field, and the maximum focused transmission increases when the emitter QPC transmission is raised to 2. The increase does not reach the maximum factor of × 2 due to the broadened collimation of the emitted beam. The trend is shared with experimental results, shown in fig 4(b).

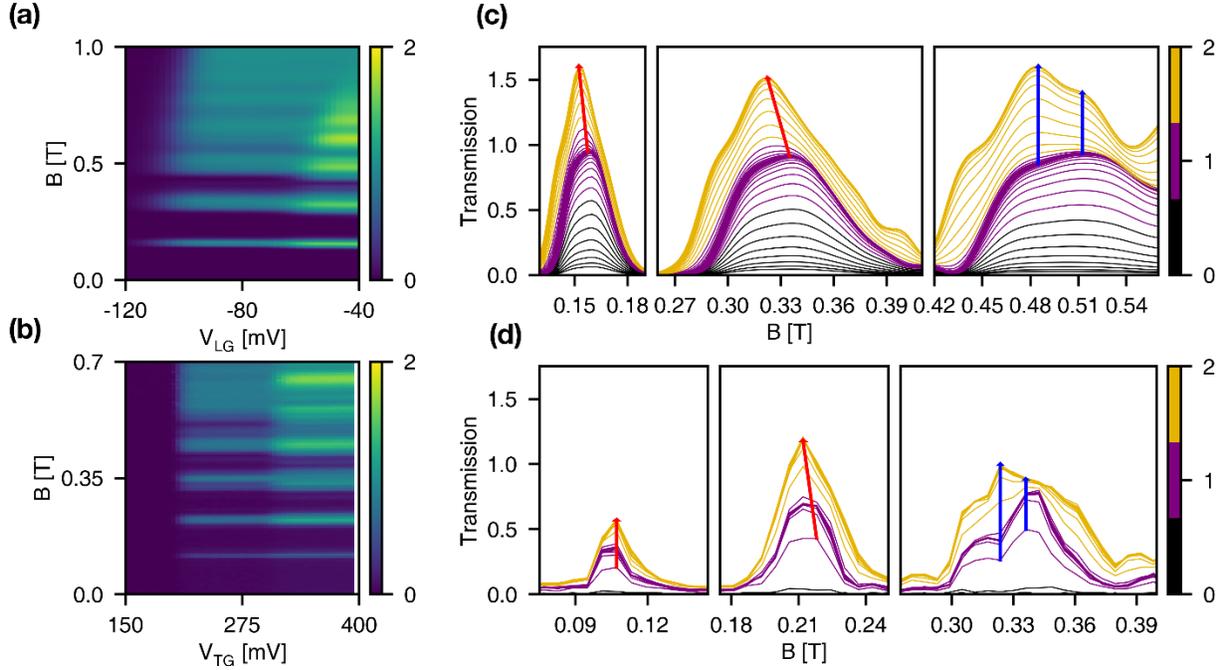

**Figure 4 Multichannel TMF.** The TMF spectrum was obtained by (a) simulation and (b) experiment, where the collector QPC transmission was $T_{col} = 2$ and the emitter QPC transmission was modulated using gate voltages (a) $V_{LG}$ or (b) $V_{TG}$. (c) The first two focusing peaks showed a steady decrease in the focusing field as the emitter QPC opened up (red). The third peak seemed to have two peak positions, each more evident at different emitter QPC transmissions (blue). (d) Experiments showed a similar trend within the available resolution.

However, a closer inspection reveals that the focusing field decreases with the rise of the emitter QPC transmission. We can consider two reasons. Increasing the QPC transmission requires raising $V_{LG}$, thereby moving the QPC center outwards. The QPC position shifts by 25 nm, which can account for a ≈ 2.5 % shift in the focusing field. This effect can be seen in the leftwards shift in the first two focusing peaks (fig. 4(c), red lines). On the other hand, the introduction of a second channel changes the collimation of particles being emitted. While the first channel exits the QPC in a straight line, the second channel exits the QPC in a pair of oblique lines [6]. Such misaligned emissions are focused at a lower field and can lower the magnetic field at which the current is maximally focused. Furthermore, if the misalignment is large enough, it can manifest as a new set of focusing peaks with a periodicity slightly less than the usual $B_0^*$ [26,38]. This effect is emphasized at higher order focusing peaks where the relative difference in $B_0^*$ is amplified. The third focusing peak exhibits such behavior, where a new peak seems to emerge as the QPC transmission is raised from $T = 1$ (Fig. 4(c), blue lines). Within the measurement resolution, this behavior is also seen in the experimental results (Fig. 4(d)).

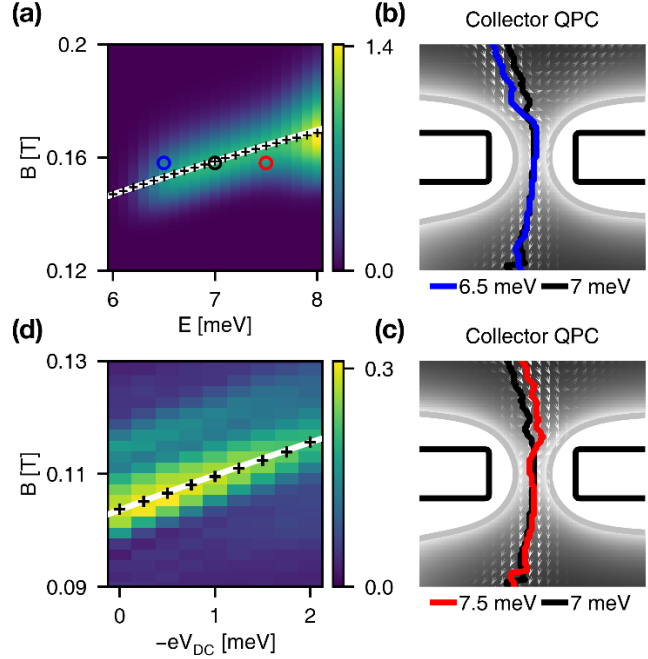

**Figure 5 Energy-modulated TMF.** (a) The simulated TMF spectrum was obtained for varying energy levels. The focusing peak position exhibited a nearly linear shift (black cross) that coincided with classical predictions (white line). The current density was examined at three different energies $E = 6.5, 7.0, 7.5$ meV at the focusing field for $E_f^* = 7$ meV. (b) When the energy was too low, the cyclotron radius was smaller, and the current fell short of being directly focused onto the collector (blue). (c) When the energy was too high, the cyclotron diameter was larger, and the current extended beyond being directly focused (red). Only at the appropriate energy was the current directly focused onto the collector ((b) and (c), black central curve). (d) The experiment peak shifts (black cross) showed good agreement with the classical predictions (white line) and, hence, the simulated results.

**Energy-modulated TMF.** Non-equilibrium studies are often difficult to interpret, partially because the results typically come in a mix between equilibrium and non-equilibrium phenomena. In KWANT, however, the equilibrium phenomena are easily obtained by modulating the energy of the calculations. In particular, the simulated results correspond to bias measurements for experiments in Fermi gas systems. This is especially relevant to TMF, where the phenomena is often employed as an energy spectrometer [12,14,21,39–45]. Figure 5(a) is the result of TMF in the energy-modulated simulations. The figure plots the first focusing peak while varying the energy by $|\epsilon| < 1$ meV about the reference energy $E_f^* = 7$ meV. Note that the QPC transmission plateau is maintained only for the range $|\epsilon| < 0.5$ meV where $\epsilon = E - E_f^*$. The focusing peak shifts (Fig.5(a), black cross) extracted using a gaussian fit (Supplementary Fig. S4) with great agreement along the classical prediction (Fig.5(a), white line):

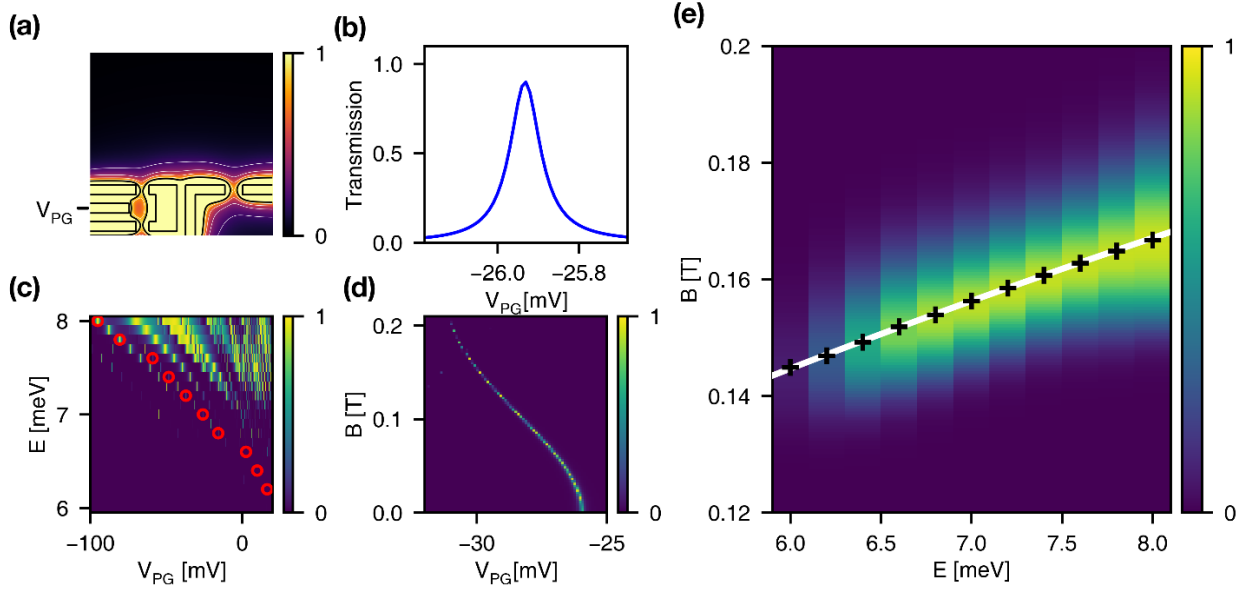

**Figure 6 Energy-modulated QD TMF.** (a) The emitter QPC of the simulated device was changed to a double-barrier potential resembling the configuration of QDs in experiment. Each white curve denotes the equipotential line with spacings $0.2 \times E_f^*$. (b) Resonance peaks in the transmission through the double-barrier potential were observed when varying the plunger gate voltage $V_{PG}$ for an appropriate set of other gate voltages. (c) At higher energies $E$, $V_{PG}$ needed to be lowered in order to match to resonance condition, as is the case in experiments. (d) A magnetic field shifted the resonance peak in a conceptually similar manner to the Fock-Darwin spectrum. (e) The simulated energy-modulated TMF spectrum exhibited peak shifts (black cross) that agreed well with classical predictions (white line) and, hence, the QPC-emitter simulations.

$$B_0^*(E_f^* + \epsilon) = B_0^*(E_f^*) \times \sqrt{1 + \epsilon/E_f^*} \tag{2}$$

derived from Eq. (1) assuming $E \propto p^2$. The classical intuition applies to the current density as well; figs. 4(b-d) are current densities for $-0.5$, $0$, and $+0.5\ meV$ at the focusing field for $E_f^*$. A particle with less energy has a smaller cyclotron radius and falls short of reaching the collector QPC (Fig. 5(b)) and vice versa (Fig. 5(c)). Only at the appropriate energy do the particles impinge upon the collector directly. In experiment, the modulated energy corresponds to the bias voltage $V_{DC}$, i.e. $\epsilon \equiv -eV_{DC}$. The good agreement of the experiment with the classical predictions (Fig. 5(d), Supplementary Fig. S5) implies that the Fermi-gas properties of real non-equilibrium measurements can be modeled by simulations.

**Energy-modulated QD TMF.** Quantum dots (QD) are often used in the study of non-equilibrium transport for their ability to filter the energy of particles [46–49]. In our simulation, the energy filtering properties of QDs can be simulated using the principle of double-barrier tunneling [24]. In the simulation, we have replaced the emitter QPC with a QD (Fig. 6(a)) and observed resonance peaks by modulating the plunger gate voltage at an appropriately tuned set of gate voltages (Fig. 6(a), $V_{PG}$ and Fig. 6(b)). When the inspected energy is heightened, $V_{PG}$ must be lowered in order to raise the resonant energy level (Fig. 6(c))—this is typically measured using the Coulomb diamond plot in experiments. The QD can be used in our energy-modulated simulation by following the resonance peaks (Fig. 6(c), red circle). Under a magnetic field, the resonance peak shifted as can be understood by the Fock-Darwin spectrum (Fig. 6(d)). The TMF spectrum can be obtained by calculating the focusing transmission along the magnetic peak shift. The process was repeated over the energy range $\pm 1$ meV about $E_F^* = 7$ meV in order to obtain the energy-modulated TMF spectrum using a QD-emitter.

The results were nearly identical to the QPC-emitter case—the focusing peaks' behavior is well-described by Eq. (2) (Fig. 6(e) and Supplementary Fig. S6). Thus, we concluded that the difference from using a QD is negligible in a Fermi-gas system. This conclusion is nontrivial for two reasons: the collimation of current leaving a QD can be different from that leaving a QPC, which would affect the TMF spectra's lineshape; also, the big difference in gate placement and the resultant potential landscape could give non-negligible perturbations to the particles' trajectory. Both of these factors important but considerably difficult to study in an experimental setting. From our simulation, however, we see that the geometric factors do not lead to major differences between the two emitter types, and the result lays a foundation upon which further experimental study can be built.

**Discussion**

Using KWANT, we show that TMF can be successfully simulated to account for the recent experiment. All parts of the experimental device with varying dimensionalities, such as the 2DEG, QPC, and QD, were simultaneously realized within each simulation and utilized minimal simplifications. Experimental TMF results were well reproduced by the numerical results, and the non-ideal behaviors observed in reality could be modeled using a disordered, random potential created by simple means. Properties of mesoscopic TMF using multiple QPC channels, a scenario in which analytic results are difficult to obtain, were also studied and showed qualitative similarities to experimental results. The current densities calculated for the single channeled TMF exhibited cyclotron motion, and its transition to skipping orbits and edge trajectories resembling quantum Hall states could be seen as well. The extension to energy-modulated cases validated the naïve use of semi-classical pictures to predict experimental focusing peak shifts and revealed the spatial modulation of the cyclotron radius for varying energy levels. Finally, the emitter QPC was exchanged with a QD, and the QD-emitter TMF simulation implied that QD usage should have minimal geometric effect on TMF for 2DEG experiments on Fermi gas systems. We believe our thorough study of simulated TMF and its comparison to experimental results proves the utility of accessible simulation power and justifies the use of such numerical approach in predicting the fundamental behavior of realistic experimental devices.

**Supplementary Materials: Numerical Reconstruction of 2D Magnetic Focusing Experiments**

Dongsung T. Park [1*], Seokyeong Lee[1*], Uhjin Kim [2], Yunchul Chung [3], Hyoungsoon Choi [1†], Hyung Kook Choi [2†]

[1] *Department of Physics, KAIST, Daejeon 34141, Republic of Korea*

[2] *Department of Physics, Jeonbuk National University, Jeonju 54896 Republic of, Korea*

[3] *Department of Physics, Pusan National University, Busan 46241, Republic of Korea*

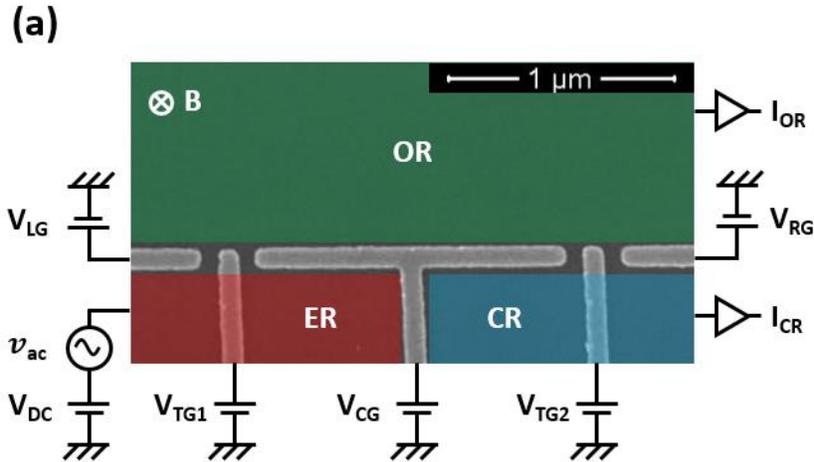

**Supplementary Figure S1. Experiment Device.** The device was fabricated on a GaAs/AlGaAs heterostructure with metallic Schottky gates lithographed on top. The two-dimensional electron gas (2DEG) residing 50 nm was divided into three reservoirs (emitter, open, and collector reservoirs; ER, OR, and CR) by placing a negative voltage $V_{gate}$ on three depletive gates (left, center, and, and right gates; LG, CG, and RG). The emitter QPC was formed between LG and CG; the collector, between CG and RG. The trench gates TG1 and TG2 were used to increase the subband separation within the QPCs [1]. A small AC voltage $v_{ac}$ was added to a DC voltage $V_{DC}$ using a bias-tee and applied to the ER. The current created were either focused onto the collector QPC and drained through the CR or reflected and drained through the OR. The focused current $I_{CR}$ and reflected current $I_{OR}$ were simultaneously measured using home-made preamplifiers connected to commercial lock-in amplifiers [2].

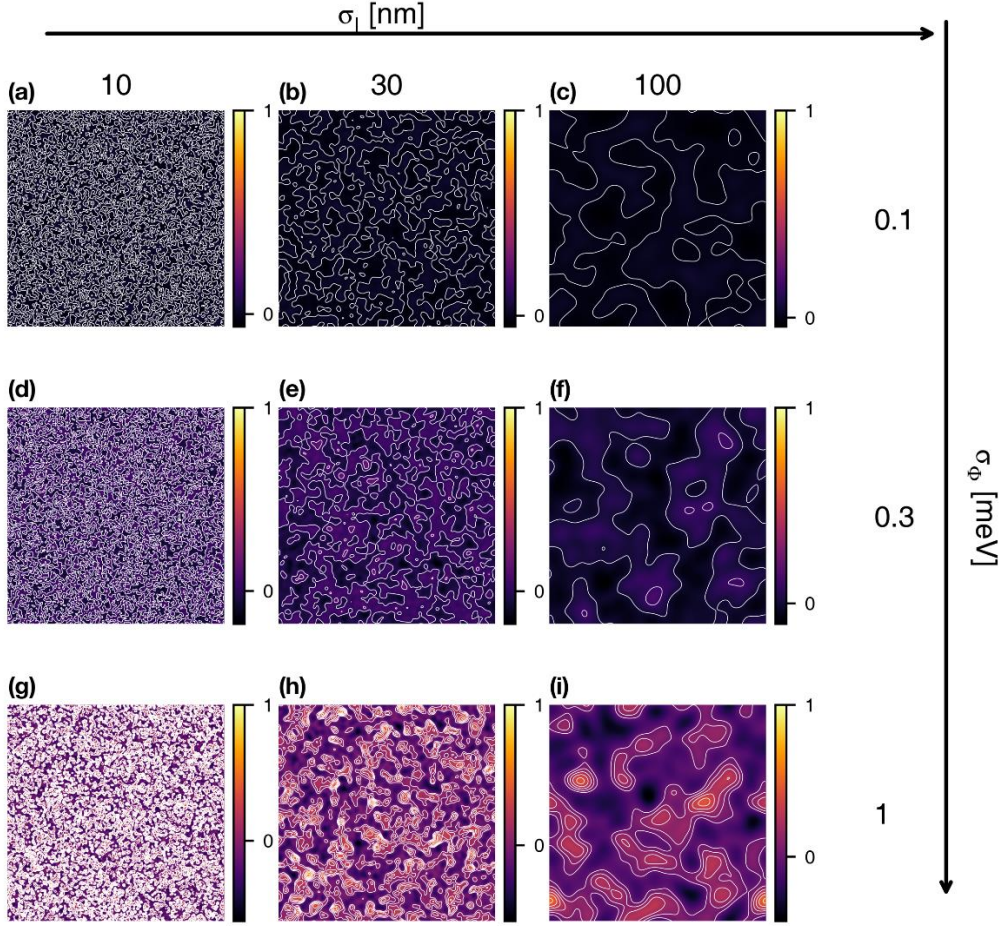

**Supplementary Figure S2. Disorder potential.** The disorder in a real sample was modeled using a random potential. A random potential field $\Phi_0$ was made following a normal potential with mean 0 and standard deviation 1, i.e. $\Phi_0(\vec{r}) \sim N(0,1) : \forall \vec{r} \in$ sites. Clearly, $\overline{\Phi_0(\vec{r})\Phi_0(\vec{r}')} = \delta^2(\vec{r} - \vec{r}')$. Then, the potential was gaussian smoothed with the kernel $K(\vec{r}\,;\sigma_l) = \exp(-(\vec{r}/\sigma_l)^2)$, i.e. $\Phi_1(\vec{r}\,;\sigma_l) = \int d^2r'\, \Phi_0(\vec{r})K(\vec{r} - \vec{r}'\,;\sigma_l)$. We note that the convolution used the periodic boundary condition, i.e. $\Phi_0(\vec{r} + \vec{R}) \coloneqq \Phi_0(\vec{r})$ where $\vec{R}$ is a vector from a corner of the simulated area to another neighboring corner. Finally, the smoothed potential was normalized and multiplied by $\sigma_\Phi$, i.e. $\Phi(\vec{r}\,;\sigma_l,\sigma_\Phi) = \Phi_1(\vec{r}\,;\sigma_l) \times \sigma_\Phi/\text{RMS}[\Phi_1(\vec{r}'\,;\sigma_l)]$ where $\text{RMS}[\Phi_1(\vec{r}'\,;\sigma_l)] = \int d^2r'\, \Phi_1(\vec{r})^2/\int d^2r'\, 1$. By construction, this gives a potential $\overline{\Phi_0(\vec{r}\,;\sigma_l,\sigma_\Phi)\Phi_0(\vec{r}'\,;\sigma_l,\sigma_\Phi)} = \sigma_\Phi^2 \exp(-(\Delta\vec{r}/2\sigma_l)^2)$, where $\Delta\vec{r} = \vec{r} - \vec{r}' \mod \vec{R}$. Subfigures (a-i) is the sample disorder potential used in the main study, where $\sigma_l$ and $\sigma_\Phi$ are varied as shown in the top and right superaxes. The contours are equipotential lines separated by an energy difference $0.1 \times E_f^* = 0.7$ meV.

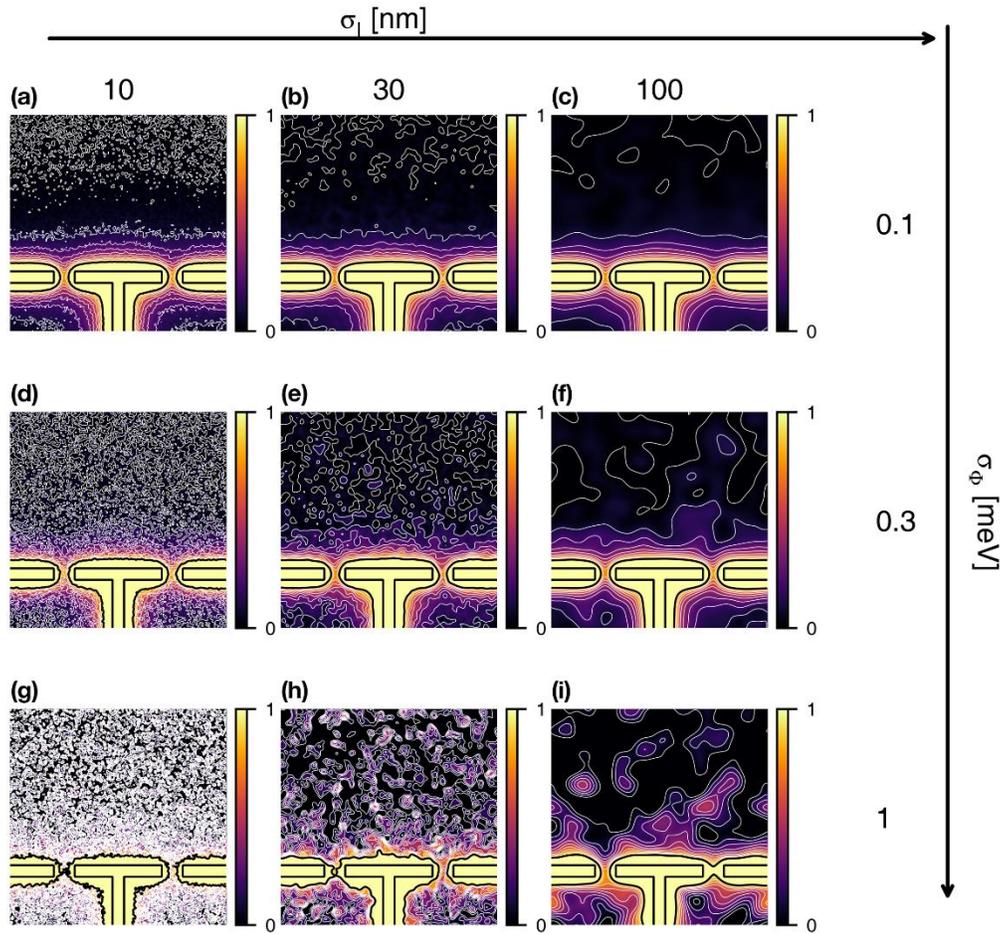

**Supplementary Figure S3. Disorder potential on Device.** The disorder potential superposed on the gate potential, i.e. $\phi[\Phi] = \phi + \Phi$, gives us the potential landscape of the sample which the electrons propagate through. Subfigures (a-i) are the sum of the gate potential and the disorder potentials shown in figure S2 (a-i) and used in the main figure 3(b-j), respectively.

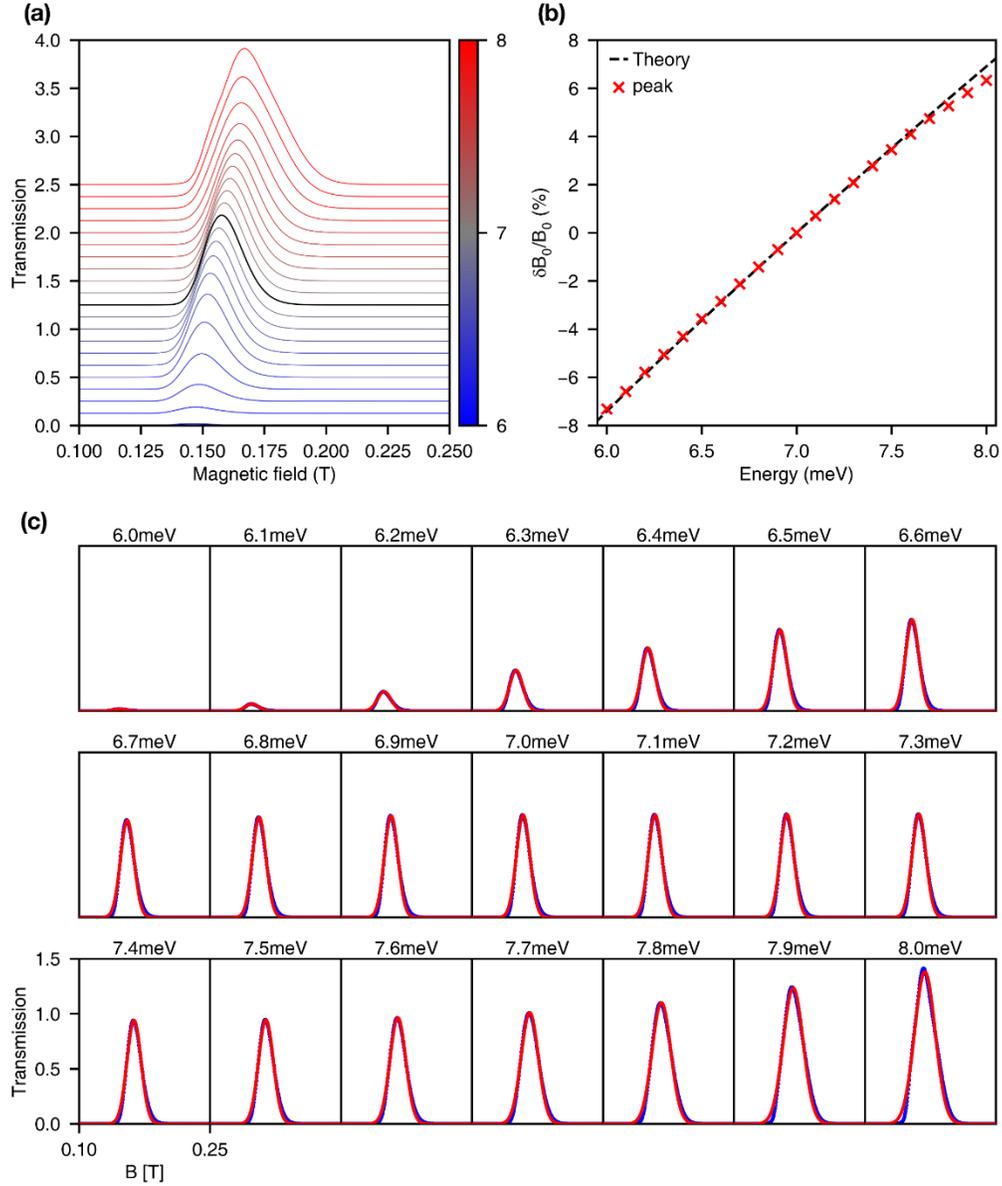

**Supplementary Figure S4. Energy-modulated TMF Simulation.** (a) The simulated TMF spectra for energies $E \in [6.0, 6.1, \ldots, 8.0]$ meV were calculated. The result for $E = 6.0$ meV is shown on the bottom in blue. The spectra for increasing energies are offset by 0.1 in an increasing red color except for the energy $E = E_f^* = 7.0$ meV which is shown in black. (b) The relative peak shift was calculated by extracting the peak position at all energies then comparing it to that from the reference energy level $E_f^* = 7.0$ meV. The peak shifts (red cross) fit well with the classical prediction (dashed line) derived in the main text, Eq. (2). At higher energies, however, a small deviation is seen, likely due to the transmission rising above $T = 1$ and incurring a multichannel effect observed in main figure 4. (c) The peak positions were extracted using a gaussian fit in order to mitigate this effect.

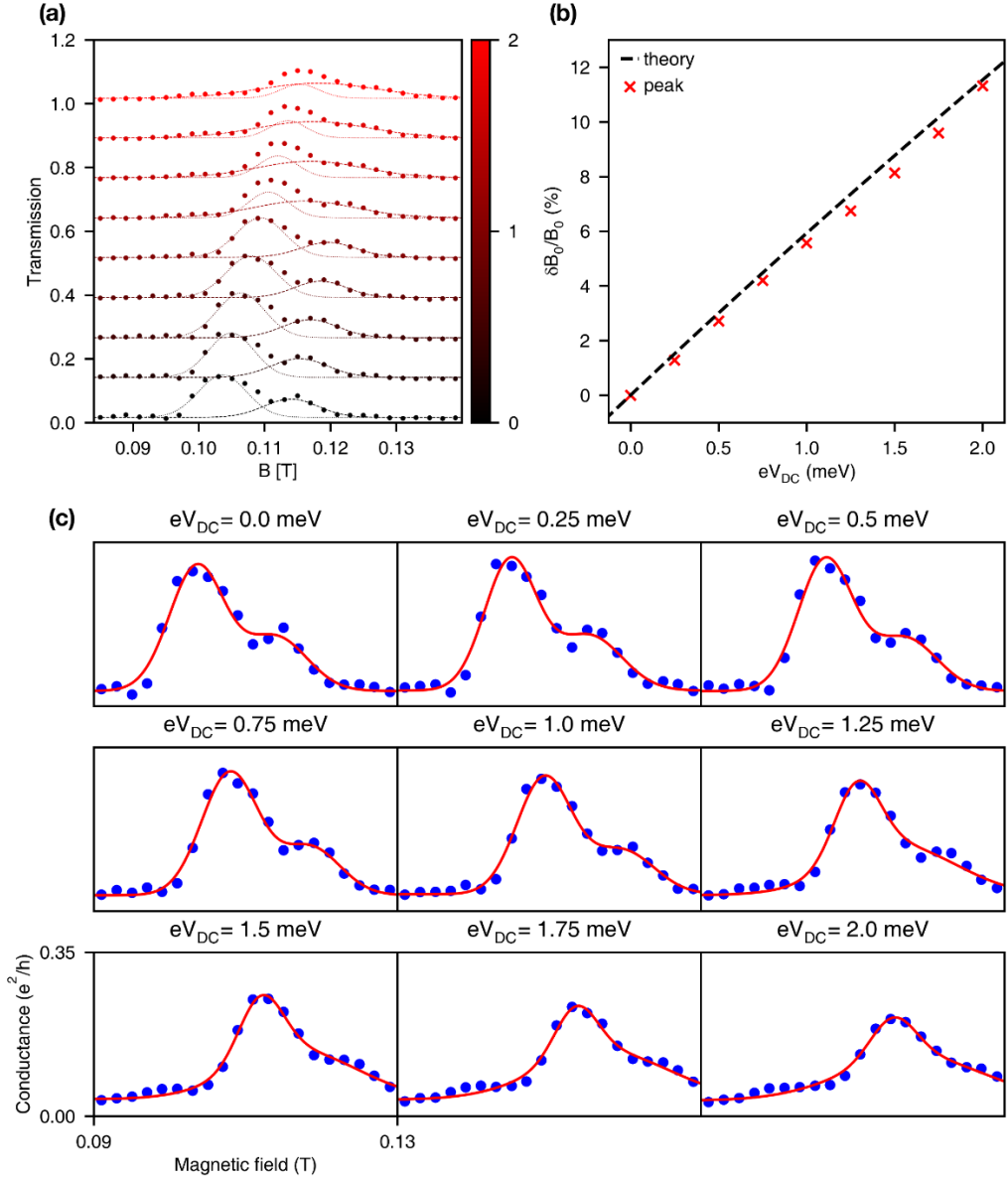

**Supplementary Figure S5. Energy-modulated TMF Experiment.** The experimental TMF spectra for biases $-eV_{DC} \in [0.00, 0.25, \ldots, 2]$ meV were measured. The figures are plotted in a manner similar to that of supplementary figure S4. Note, however, that a double-gaussian fit was used to account for the unidentified satellite peak occurring at a higher magnetic field. A satellite peak may emerge due to the disorder creating a branched current flow leaving the emitter that follows a perturbed cyclotron orbit with a slightly extended diameter. Nevertheless, the larger gaussian line was easily identified for $-eV_{DC} \leq$ 1.5 meV and extrapolated to $-eV_{DC} \leq 2.0$ meV. The identified peak positions showed good agreement with the classical predictions as well.

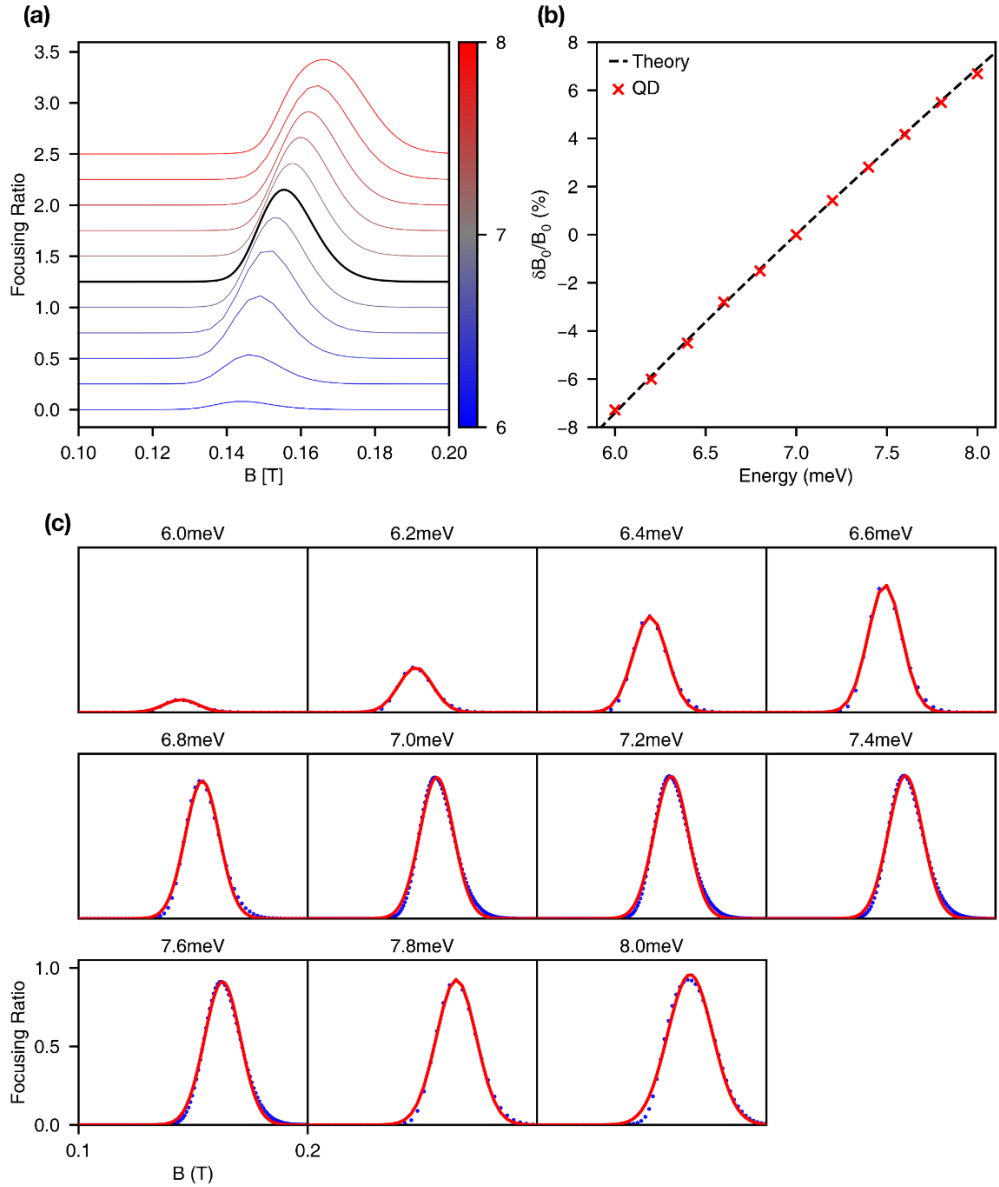

**Supplementary Figure S6. Energy-modulated QD TMF Simulation.** The simulated QD TMF spectra for energies $E \in [6.0, 6.2, \ldots, 8.0]$ meV were calculated. The figures are plotted in a manner similar to that of supplementary figures S4 and S5. Note, however, that the spectra are shown not by the transmission but the focusing ratio, i.e. $T_{CR}/(T_{CR} + T_{OR})$. This accounts for two factors: the changes in the double-barrier transmission due to the magnetic field; and the simulation not occurring at the exact $V_{PG}$ at which the resonance occurs because of resolution limitations. However, the transmission through a double-barrier potential is $T \leq 1$, and the focusing ratio should scale exactly with the TMF spectra given by the transmission at the double-barrier resonance peaks.